# Labor Reforms in Rajasthan: A boon or a bane?

Diti Goswami[1]    Sourabh Paul[2]

We examine the impact of labour law deregulations in the Indian state of Rajasthan on plant employment and productivity. In 2014, after a long time, Rajasthan was the first Indian state that introduced labour reforms in the Industrial Disputes Act (1947), the Factories Act (1948), the Contract Labour (Regulation and Abolition) Act (1970), and the Apprentices Act (1961). Exploiting this unique quasi-natural experiment, we apply a difference-in-difference framework using the Annual Survey of Industries panel data of manufacturing plants. Our results show that reforms had an unintended consequence of the decline in labour use. Also, worryingly, the flexibility resulted in the disproportionate decline in the directly employed worker. Evidence suggests that the reforms positively impacted the plants' value-added and productivity. The strength of these effects varies depending on the underlying industry and reform structure. These findings prove robust to a set of specifications.

[1] Indian Institute of Technology Delhi. Email: Diti.Goswami@hss.iitd.ac.in
[2] Indian Institute of Technology Delhi

# Introduction

Internationally, Indian labor laws are considered rigid and complex. In this vein, recently, the Indian government passed three major labor code bills by the Parliament: The Industrial Relations Code Bill, 2020; the Code on Social Security Bill, 2020; and the Occupational Safety, Health, and Working Conditions Code Bill, 2020 along with the Code on Wages Bill enacted in 2019. After a long time, the government introduced these new laws to reduce complexities, bring more transparency and accountability, and help both employers and workers. These reforms in the labor laws with a high degree of both political and public interest started back in 2014, with Rajasthan being the first Indian state to deregulate the labor laws in four major Acts (Government of India, 2018).

The reforms in India's labor laws resulted from a rigorous debate. One strand of literature argued the restrictive labor laws hurt the firms by forcing them to remain small and use contractual workers or capital-intensive technologies (Amirapu and Gechter 2020; Hasan, Mehta, and Sundaram 2020; Hasan, Kapoor, Mehta, and Sundaram 2017; Ahluwalia, Hasan, Kapoor, and Panagariya 2018). In contrast, the other strand of literature opined the labor laws could not be held responsible for the Indian economy's sluggish growth (D'Souza 2010; Chatterjee and Kanbur 2015; Roychowdhury 2014; Roy, Dubey, and Ramaiah 2020; Deakin and Haldar 2015). These two strands of opinion on the Indian labor market flexibility differ in various theoretical understanding, methodological details, and empirical ground.

Believing that the strict labor laws are detrimental to the Indian economy, the Indian government started relaxing some of the major Indian labor laws both at the national and sub-national levels (Government of India, 2018). These important reforms in the Indian labor market require careful independent evaluation. The labor reforms in 2014 in the Indian state of Rajasthan provides a natural experiment to understand the impact of such reforms. In particular, in this paper, we find the causal impact of the labor law reforms in Rajasthan on the plant's employment and performances. Rajasthan deregulated the labor laws in the Industrial Disputes Act (1947), the Factories Act (1948), the Contract Labor (Regulation and Abolition) Act (1970), and the Apprentices Act (1961). These reforms in the labor laws in Rajasthan allowed us to utilize a quasi-natural experimental research design. We use a difference-in-difference specification to the establishment-level Annual Survey of Industries (ASI) longitudinal data from 2011-12 to 2016-17 to examine the effects of Rajasthan's pro-employer reforms on employees, direct and contractual workers, capital, inputs, gross value added (GVA), total factor productivity (TFP), profits, and workers' emoluments. As the existing literature does not provide any clear-cut opinion of the benefits of the labor market flexibility in the Indian economy, it is not surprising that the empirical outcome of the newly introduced labor reforms on plant employment and performances are ambiguous. Our work contributes to the existing literature by providing evidence on whether the newly introduced flexibility in the labor laws is gainful or not. The novelty of our work is that understanding the impact of deregulations in the Indian state of Rajasthan will help predict the implications of the recent national amendments in labor laws that affect nearly 425 million Indian working-age population. Thus our study on the causal effect of deregulations in the labor laws generates important policy-relevant insights.

Our article is not the first to examine the effect of the amendments in labor laws in Rajasthan. Most notably, the Economic Survey 2018-19 of India (Government of India, 2018) exploited the ASI data from 2011 to 2017 to analyze the reforms in Rajasthan. Our work is distinguished from the Economic Survey (Government of India, 2018) by the aggregation level and the outcome variables. In particular, the Economic Survey finds a macro impact (extensive margin) of the reforms on the total manufacturing employment and other variables like the number of factories with greater than 100 workers, total output, total wages, output per factory, and wages per factory using 146 observations at the state level. In contrast, our article finds the micro impact (intensive margin) of the reform on the plant's employment and performances. Provided the heterogeneity among the manufacturing plants at the aggregate level, our study provides a deeper understanding of the reforms (Mathew 2017; Padmaja and Sasidharan 2016).

Our empirical analysis provides evidence that the reforms had an unintended consequence of the decline in labor use. The implications regarding employment change are similar to Roy, Dubey, and Ramaiah (2020), Deakin and Haldar (2015), Roychowdhury (2019a); Chatterjee and Kanbur (2015), D'Souza (2010), Kapoor (2014), Chandru (2014) in the sense that higher flexibility causes weaker employment growth. Also, worryingly, the increased flexibility resulted in the decline in directly employed workers. We find the plants that directly fall under the Industrial Disputes Act (1947) reforms experience an expansion in labor compared to the plants that are not under the direct ambit of the Industrial Disputes Act (1947) reforms. Moreover, our data show that the plants that fall directly under the Contract Labor (Regulation and Abolition) Act (1970) reforms experience greater use of contractual workers than the plants that are likely to be unaffected. We also find reforms to cause the plants in the labor-intensive industries to restructure their production mix by reducing their labor use. In contrast, the newly introduced labor laws' flexibility caused the plants in export-oriented industries to use more contractual workers. Regarding the impact of the reforms on plant inputs and performances, we find the reforms to positively impact the plants' value added and productivity. The parallel test result indicates no change in the plant outcomes before the reforms in the treatment and control states. Furthermore, the authors find that the reforms did not cause new manufacturing plants in Rajasthan.

## Literature Review

Indian labor laws have been the focus of many debates. One strand of literature argues against labor protection on the grounds of strict labor laws; i) directly ii) indirectly reducing the economy's efficiency, and iii) increasing labor substitution with capital or contractual workers. In contrast, the other literature opines that the labor laws could not be held responsible for the Indian economy's sluggish growth.

### *i) Direct adverse impact of the strict labor regulation on plant performances and economic outcomes*

One of the influential studies on the detrimental effects of labor protection on the Indian economy is Besley and Burgess (2004). They find the Indian states' pro-worker regulations to cause lower output, employment, productivity, and investment in the formal manufacturing sector. Another similar study by Ahsan and Pagés (2009), find an adverse impact of employment protection and cost of dispute resolution on employment and output. Moreover, this adverse impact is more for the states and time-frame, where the cost to resolve a dispute is high. Workers do not benefit from these protections as the authors do not find an increase in labor share or wage bill. Bhattacharya, Narayan, Popp, and Rath (2011) find the rigid labor market in India to hinder the multinationals from operating in the labor-intensive production process compared to countries like China and the Philippines. Also, Lee (2019) finds a lack of labor demand in rigid labor markets in India. This strand of literature opines labor market reforms to arrest the detrimental effects of high labor cost and rigidity. The reforms will improve wage share, control the increase of informal employment, and increase aggregate productivity (Dougherty 2009). Amin (2009) analyze the impact of labor laws on the employment of 1948 retail stores in India. He reports that 27% of the stores find labor regulation as a hinder to their business activities. He finds that the labor reforms will increase employment by 22% for an average store. Further he finds the strict labor laws to increase labor costs, resulting in firms substituting labor with the computer (Amin 2007). Pro-worker legislation or labor unrest also adversely impact the location choice and investments (Menon and Sanyal 2007; Sanyal and Menon 2005). Dougherty, Robles, and Krishna (2011) find that strict labor regulations are likely to harm industries with high labor intensity or high sales volatility.

They estimate that firms experience a 14% higher TFP in labor-intensive industries and the states with the flexible labor market than the firms in the labor-intensive but rigid labor market. Similarly, they experience 11% higher TFP in a pro-employer state than in the pro-worker state for the volatile industries. One recent study by Hasan, Mehta, and Sundaram (2020) finds that the rigid labor regulations to adversely impact the exporters by reducing the output. The rigid firing restriction also reduces the firm's employment responses to temporary shocks (Adhvaryu, Chari, and Sharma 2013). Another impact of rigid labor laws is an increase in corruption. A recent study by Amirapu and Gechter (2020) estimates that regulations increase firms' labor costs by around 35%, which increases the possibility of corruption[3].

## ii) Indirect adverse impact of the strict labor regulation on plant performances and economic outcomes

Aghion, Burgess, Redding, and Zilibotti (2008) find the dismantling of License Raj helps the industries in pro-employer states to grow more quickly than in pro-worker states. Mitra and Ural (2008) find the positive impact of trade liberalization on productivity more pronounced for states with a flexible labor market. They also find that trade liberalization helps the export-oriented industries in states that have flexible labor laws. Labor demand elasticity is also higher with trade liberalization for the states with flexible labor markets (Hasan, Mitra, and Ramaswamy 2007). Labor regulations are not only limited to generate gains from trade liberalization and are also crucial in the firm size distribution. Larger sized firms are prevalent in the states with flexible labor regulation. The prevalence of large-sized firms in flexible states is more prominent among the firms that started production after 1982, when labor laws were tightened (Hasan and Jandoc 2012). Thus, Hasan and Jandoc (2012) claim labor regulations to affect the firm's size adversely. Hasan, Kapoor, Mehta, and Sundaram (2017) emphasize that even if India is one of the largest producers and exporter of apparel, the sector is still to operate at its potential. They point this incapability to the labor regulations that cause the firms to operate at scales which are insufficient to use the advanced techniques. Another recent study by Ahluwalia, Hasan, Kapoor, and Panagariya (2018) analyzes the impact of labor regulations on employment and wages. They use the 2005 abolition of the quota restrictions on the export of apparel and textile products from developing to developed countries and the variation in the labor regulations across the Indian state as a natural experiment to find the effect of labor regulation. They find significant benefits in employment and wages post 2005 in the apparel and textile industries in states with flexible labor laws.

## iii) Strict labor regulation cause substitution of labor with capital and temporary employment

Hasan, Mitra, and Sundaram (2013b) find the capital intensity higher for India's firms than other countries with the same level of economic development or factor endowments. They find strict labor regulations as one of the primary reasons for the high capital intensity. The rigid labor laws do not help trade gains based on factor abundance comparative advantage (Hasan, Mitra, and Sundaram 2013a). Hasan, Mehta, and Sundaram (2020) find that producers in pro-worker states replace labor with capital. Firms use contractual or fixed-term workers for many reasons (Singh, Das, Abhishek, and Kukreja 2019). Some of the reasons are; to reduce the high labor cost (Sapkal 2016; Basu, Chau, and Soundararajan 2018), reduce the bargaining power of the permanent workers (Saha, Sen, and Maiti 2013), stay away from the legal establishment size threshold of 100 workers (Ramaswamy 2013a; Ramaswamy 2013b), increase flexibility as the employers are free to hire and fire the contract workers (Srivastava 2016), deal with temporary shocks (Chaurey 2015), and many more. Ramaswamy (2013a) finds the strict labor laws to cause higher intensity of contract workers for the size group of 55-99 workers and in labor-intensive inflexible states. A regional case study by Barnes, Das, and Pratap (2015) in North India's automotive components production shows how a regional contract labor system has helped the employers to keep the wages low, enjoy more flexibility, skip the burden of monitoring and controlling the workers, and weaken the labor rights. However, the use of a contract worker has its demerits. As it is an 'incomplete contract', the workers' underinvest in specific skills (Singh, Das, Abhishek, and Kukreja 2019; Singh, Das, Kukreja, and Abhishek 2017).

---

[3] The adverse impact of the strict labor laws have been discussed for other countries also; like Yoo and Kang (2012) and Baek and Park (2018) for South Korea, Ingham and Ingham (2011) for Poland, Bossler and Gerner (2020) for Germany .

*Criticism of the view that Indian labor regulations harm the Indian economy*

The regulations on job security do not negatively affect as (D'Souza 2010) as firms transform the work practices and make it flexible through non-compliance or weak enforcement of laws (Chatterjee and Kanbur 2015). Badigannavar and Kelly (2012) finds that even a pro-worker state like Maharashtra provides weak protection to the formal sector workers and the labor unions. A recent study by Roy, Dubey, and Ramaiah (2020) finds no evidence of the spatial variation in labor regulations flexibility in explaining employment growth variation. They find that higher flexibility associates with weaker employment growth. In a similar vein, Roychowdhury (2019a) and Roychowdhury (2019b) explain that the labor laws cannot be held responsible for the employment stagnation in India's organized manufacturing as it applies to less than 35% of aggregate employment. He further finds that the worker's bargaining power is declining in Indian manufacturing. A study on Gujarat's deregulatory reforms by Deakin and Haldar (2015) proposes very little evidence linking law deregulation to growth. In response to the belief that employment protection legislation restricts employment adjustment from demand shock, Sofi and Kunroo (2018) find no evidence from 2000-01 to 2011-12. Moreover, Rodgers and Menon (2013) find that employment adjustment and dispute settlement restrictions cause higher job quality for women.

One of the most influential studies by Besley and Burgess (2004) has been criticized on various conceptual and measurement issues, coding errors, methodological problems, failure to replicate the findings (Bhattacharjea 2006; Bhattacharjea 2009; Jha and Golder 2008; Storm 2019; D'Souza 2010) and, difficulties in the enforcement of the labor laws in India (Fagernäs 2010). Some scholars find the labor laws changes are endogenous to several other economic factors and do not explicitly determine economic indicators (Dutta Roy 2004; Deakin, and Sarkar 2011). Another study that has been severely criticized is Basu, Fields, and Debgupta (2009), which find flexible labor laws beneficial to workers' wages and employment. Roychowdhury (2014) examine their theoretical argument and find their policy conclusion to be unsustainable.

Thus, the impact of strict labor laws on the Indian economy is inconclusive. One group of scholars advocate relaxation in labor laws, while the other group claims does not. The differences in opinion stem from both analytical as well as methodological understanding. In this context, the Indian government deregulated the labor laws in recent years, believing that the pre-existing labor laws are detrimental (Government of India, 2018). We analyze the impact of these recent relaxations in labor laws on plant employment and performances. Thus, this paper adds to the existing literature by finding whether the recent relaxations in the labor laws that one group of scholars have been advocating over the years have been gainful or not. This paper contributes to the literature on the Indian institutional reform effects, plant employment and performances in Indian manufacturing, quasi-natural experiment, and on the recent developments in the Indian labor market.

## Background

Under the Constitution of India, labor is a subject in the concurrent list where both the Central and the state governments are capable of enacting legislation. None of the Indian states adopted labor reforms previous to 2014 (Government of India, 2018). After a long time, Rajasthan was the first state that introduced labor reforms in four majors Acts: The Industrial Disputes Act (1947), The Factories Act

(1948), The Contract Labor (Regulation and Abolition) Act (1970), and the Apprentices Act (1961) in 2014. Table 1 describes the amendments in each of these Acts.

**Table 1: Labor Reforms in Rajasthan**

|  | Major Amendments |
|---|---|
| The Industrial Disputes Act (1947) | i. Government approval is not required for companies with 300 or fewer workers to shut down or retrench workers. The earlier limit was 100 workers.<br>ii. The membership requirement to form a union has increased from 15% to 30% of the total workmen.<br>iii. The time limit for any worker to object has been reduced to three years from an indefinite period. |
| The Factories Act (1948) | i. The threshold limit increased from 10 or more workers with the power to 20 or more workers with power, and 20 or more workers without power to 40 or more workers without power.<br>ii. Any complaint against the employer about the violation of this Act will not receive cognizance by a court without prior permission from the state government. |
| The Contract Labor (Regulation and Abolition) Act (1970) | i. Applicable to establishments that employ 50 or more workers on contract against the former 20 or more workers. |
| The Apprentices Act (1961) | i. Apprentice's stipend will be no less than the minimum wage.<br>ii. Government to bear part of the costs of apprentice training in order to encourage skilling. |

All these reforms were pro-employer. These deregulations in the labor laws provide an interesting setup to examine the impact of the reforms on plant's employment and performances.

Diluting of the labor laws can have an ambiguous impact on the plant's employment, performances, and inputs. The flexibility in labor laws can increase plant employment as employers prefer labor over other production factors. As the reform reduces the hiring and firing cost, employers can adjust the laborers according to their requirements and prefer cheap labor. The labor's bargaining power also reduces, and this might act as a further boost to increase employment. These reforms can also increase contractual workers' use because of their added advantage (Drager and Marx 2017; Kuroki 2012). However, these pro-employer reforms can also cause employment to decline as employers get the authority to shed workers. The lack of powerful labor unions further makes the dismissal process easy (Watanabe 2018; Roychowdhury 2019a; Roychowdhury 2019b). As the labor cost declines with the reforms, plants might increase capital investments to complement the labor. Also, flexible labor laws result in less costly bank loans as the borrower's default risk declines due to the increased flexibility to adjust labor (Alimov 2015). On the contrary, the low cost of labor can cause employers to substitute labor for capital (Hasan, Mitra, and Sundaram 2013a; Hasan, Mitra, and Sundaram 2013b; Hasan, Mehta, and Sundaram 2020). Plant productivity may increase because employers can adjust laborers and lay off unproductive workers resulting in the most productive skill matches (Caballero, Cowan, Engel, Micco 2013; Maida and Tealdi 2020). The worker's effort can also increase because of the fear of dismissal (Bradley, Green, and Leeves 2014). On the contrary, productivity can decrease as low job security might cause the workers to invest less in plant-specific human capital value addition (Acharya, Baghai, and Subramanian 2013), discourage the workers from providing effort, and high wage inequality among the workers (Shimizutani, and Yokoyama 2009; Silva, Martins and Lopes 2018). Thus, the effect of the labor laws reforms on plant employment, inputs, and performances warrants an empirical examination.

# Empirical Methodology

We use difference-in-difference (DID) framework to compare the plant outcomes before and after the reforms in the treatment and the control state. To the best of our knowledge, we find no other policy or reforms implemented in Rajasthan in 2014 that impacted plant-level outcomes differentially more or less than in the control state. This will help us identify the treatment effect of labor reforms in Rajasthan.

Ideally, we would like to compare the plants in Rajasthan with an observationally similar control group. We choose the establishments in Punjab as a control group because of the following reasons: i) Punjab is a neighboring state of Rajasthan and have similar characteristics in various aspects ii) Among the other neighboring states of Rajasthan, i.e., Gujarat, Madhya Pradesh, Uttar Pradesh, and Haryana had also started to think of introducing flexibility in labor laws around 2014 mostly because of the similar political affiliation in those states (ruling party). Thus, we would not get correct estimates if we use these states as a control group. iii) Punjab and Rajasthan experience a similar degree of flexibility in labor restrictions before the reform (Government of India, 2018). Furthermore, as a robustness check, we choose establishments from other states as a control group.

The identifying assumption of the DID estimator is that the treatment and the control group should have similar trends before the reform. In the subsequent section, we show that Rajasthan and Punjab have similar trends in pre-2014. We estimate the following plant level regression specification in Equation (l) to find the impact of the amendments in labor laws on various plant outcomes.

$$Y_{ijst} = \beta_0\, Treat_i + \beta_1\, Post_t + \beta_2\, Treat_i * Post_t + X_{ijt} + \kappa_i + \gamma_t + \delta_j t + \epsilon_{ijst} \quad (1)$$

$i,j,s,t$ in $Y_{ijst} = \beta_0\, Treat_i + \beta_1\, Post_t + \beta_2\, Treat_i * Post_t + X_{ijt} + \kappa_i + \gamma_t + \delta_j t + \epsilon_{ijst}$ ( 1 index plants, industry (2-digit), state and year. $Y_{ijst}$ represents plant level outcomes like employers, direct workers, contractual workers, capital, inputs, GVA, TFP, profit, and emoluments. $Post_t$ is an indicator variable that takes the value of 1 for the years after the amendments (2014-15 to 2016-17) and 0 otherwise. $Treat_i$ is an indicator variable that takes the value of 1 if the plant is in the treated state Rajasthan and 0 if the plant belongs to Punjab. $\kappa_i$ is the plant fixed effect that controls for any time-invariant unobserved heterogeneity. $\gamma_t$ is the year fixed effect that controls the year specific unobserved changes. We should keep in mind that $Post_t$ will be completely absorbed by the year fixed effects whereas, $Treat_i$ will be completely absorbed by the plant fixed effects. $X_{ijt}$ are the controls, namely, age of the plant, percentage of the output that the plant export, import dummy trend, GVA, capital, inputs, profit, emoluments, and workers[4]. $\delta_j t$ represents industry trends and $\epsilon_{ijst}$ is the stochastic error term. We cluster standard errors at the state levels. The coefficient of the interaction of $Treat_i$ and $Post_t$, $\beta_2$ is the coefficient of our interest that captures the causal impact of the labor laws reform on the plant outcome variables. We consider the entrants, incumbents, and exiters during the sample period.

The reforms in the labor laws may be more pronounced for the "affected plants". Affected plants are those that are most likely to be affected by the labor laws reforms. We identify affected plants in two ways; i) Plants that fall directly under the Industrial Disputes Act (1947) reforms, and ii) Plants that fall directly under the Contract Labor (Regulation and Abolition) Act (1970) reform, in the pre-treatment period. To find the impact of the reforms on the affected plants, we use triple difference and estimate the following regression specification:

$$Y_{ijst} = \beta_0\, Treat_i + \beta_1\, Post_t + \beta_2\, Affected_i + \beta_3\, Treat_i * Post_t + \beta_4\, Post_t * Affected_i + \beta_5\, Treat_i * Affected_i + \beta_6\, Treat_i * Post_t * Affected_i + \kappa_i + \gamma_t + \delta_j t + \varepsilon_{ijst}$$
(1)

---
[4] We exclude the variable as a control that is the dependent variable for a particular regression specification.

The coefficient, $\beta_6$ captures whether the affected and the non-affected plants responded differently after the reform to before in Rajasthan compared to Punjab. A significant $\beta_6$ indicates that the law changes were effective in impacting those plants that were intended to.

We further analyze the impact of the reform on heterogeneous industry categories. Plants in labor-intensive industries or export-oriented industries are more likely to be impacted by increased flexibility in labor laws. Many studies have found the strict labor regulations to affect the exporters and the labor-intensive industries adversely (Hasan, Mehta, and Sundaram 2020; Mitra and Ural 2008; Saha, Sen, and Maiti 2013; Dougherty, Robles, and Krishna 2011; Hasan, Mitra, and Sundaram 2013a; Ramaswamy 2013a). Therefore, in a similar vein, the increase in flexibility in labor laws should be differentially larger in these types of industries. To test this, we estimate the following regression specification:

$$Y_{ijst} = \beta_0\, Treat_i + \beta_1\, Post_t + \beta_2\, IndustryType_i + \beta_3\, Treat_i * Post_t + \beta_4\, Post_t * IndustryType + \beta_5\, Treat_i * IndustryType_i + \beta_6\, Treat_i * Post_t * IndustryType_i + \kappa_i + \gamma_t + \delta_j t + \varepsilon_{ijst}$$
(2)

$\beta_6$ finds the heterogeneous impact of the changes in labor laws on plant outcomes. $IndustryType_i$ is an indicator variable that takes a value of 1 if a plant is in labor-intensive/export-oriented industries and otherwise zero.

## Data

We use the plant longitudinal data of the Indian manufacturing from the Annual Survey of Industries (ASI) provided by the Ministry of Statistics and Programme Implementation, Government of India. The ASI is a nationally representative survey of plants/establishments that are registered under The Factories Act, 1949. The Factories Act, 1949, is important legislation that regulates India's manufacturing activities for and all establishments that employ 10 or more workers (with electricity) or 20 or more workers (without electricity). Since the labor reforms pertain to the formal sector, our data set covers the formal manufacturing in India. The establishments in ASI data are divided into a census sector and a survey sector. The plants with more than 100 workers or that file joint returns in the ASI survey or are situated in some industrially backward states like Manipur, Meghalaya, Nagaland, Tripura, Andaman, and Nicobar Island are surveyed every year and hence are called census sector. Plants that do not fall in the census sector are randomly sampled using a systematic circular sampling technique within each state*Industry*Sector*4 digit stratum and form the survey sector. We use the information from both the census and the sample sector for manufacturing plants. Furthermore, as a robustness check, we restrict our sample with the census sector's establishments and use a balanced panel. Table 2 presents the number of observations by the census and the sample sector in the treatment and control group.

**Table 2: Number of Observations**

|  | Rajasthan(Treatment) | | Punjab(Control) | |
| --- | --- | --- | --- | --- |
|  | Census | Sample | Census | Sample |
| 2011-12 | 638(34.71) | 1200 | 731(27.02) | 1974 |
| 2012-13 | 1726(72.12) | 667 | 1568(65.33) | 832 |
| 2013-14 | 1462(67.03) | 719 | 1392(58.14) | 1002 |
| 2014-15 | 683(29.21) | 1655 | 684(27.15) | 1835 |
| 2015-16 | 1810(66.27) | 921 | 1200(46.11) | 1402 |
| 2016-17 | 1608(60.47) | 1051 | 1215(43.37) | 1586 |

Source: Authors' calculation based on ASI data. Note: The bracketed number is the observation percentage in the census sector in that particular group year.

In this study, we utilize the ASI dataset from 2011-12 to 2016-17. The reference period of the ASI data is a fiscal year between April to March. We use plant-level information on employees, direct and contractual labor, capital, inputs, profits, emoluments, GVA, and TFP. Capital is measured as the average of fixed capital's net book value at the beginning and the end of the fiscal year. The labor input is measured as either the average number of person worked. The average number of person worked is the ratio of total person-days to the number of working days. We estimate TFP using the methodologies proposed by Woolridge (2009) and Levinsohn and Petrin (2003) as TFP (method 1) and TFP (method 2), respectively. The procedure for estimating the TFP is presented in the Appendix. GVA is deflated by the suitable wholesale price index (WPI) by groups using 2005 as the base year. Matching the detailed categories of WPI with the 2-digit industry classification was impossible due to data limitations. However, a close and mindful comparison of the groups was undertaken to choose appropriate price deflators. Fixed capital is deflated using WPI for machinery and equipment. The consumer price index (CPI) of rural laborers and industrial workers is used as a deflator for workers' emoluments. We classify an industry as labor intensive if the capital intensity[5] is below the median of the total manufacturing (Kapoor 2015). To find the export-oriented industries, we use the 2 digit industry trade information from U.N. Comtrade and calculate the value of T, where $T = \frac{M-X}{Q-X+M}$, M is import, X is export, and Q is production. If T's values is negative, then that particular industry is export-oriented (Erlat 2000; Krueger, Lary, Monson, and Akrasanee 1981). We use the output data (Q) from the United Nations Industrial Development Organization. We use the information from the pre-treatment years (2011-12 to 2013-14) to categorize the industries. Table 3 summarises the data used in our analysis for the treatment and the control groups pre and post-reform.

**Table 3: Summary Statistics**

| | Rajasthan (Treatment) | | Punjab (Control) | |
|---|---|---|---|---|
| | Pre Treatment | Post Treatment | Pre Treatment | Post Treatment |
| Observations | 6,412 | 7,728 | 7,499 | 7,922 |
| ***Panel A: Plant Employment*** | | | | |
| Log Employees | 3.120 [1.214] | 0.700 [0.150] | 3.134 [1.165] | 0.004 [0.124] |
| Contractual to Total Workers Ratio | 0.213 [0.382] | 0.195 [0.368] | 0.294 [0.413] | 0.322 [0.430 ] |
| Direct to Total Workers Ratio | 0.923 [0.212] | 0.882 [0.284] | 0.858 [0.276] | 0.785 [0.361] |
| Log Contractual Workers | 3.433 [1.210] | 3.590 [1.306] | 3.043 [1.143] | 3.111 [1.108] |
| Log Direct Workers | 2.579 [1.148] | 2.541 [1.191] | 2.504 [1.244] | 2.459 [1.301] |
| ***Panel B: Plant Performances and Inputs*** | | | | |
| Log Capital | 10.220 [3.017] | 7.549 [1.184] | 9.845 [2.229] | 9.355 [1.482] |

---

[5] Capital intensity is the ratio of fixed capital to total persons engaged.

| | | | | |
|---|---|---|---|---|
| Log Inputs | 11.811 [2.171] | 11.819 [2.242] | 11.324 [2.011] | 4.602 [1.607] |
| Log GVA | 10.485 [1.612] | 10.902 [1.838] | 10.165 [1.383] | 10.568 [1.554] |
| Log TFP (Method 1) | 7.364 [0.987] | 7.735 [1.159] | 7.122 [0.771] | 7.487 [0.914] |
| Log TFP (Method 2) | 7.476 [1.000] | 7.850 [1.173] | 7.227 [0.782] | 7.593 [0.926] |
| Log Profit | 8.720 [2.011] | 9.705 [2.338] | 8.298 [1.602] | 9.184 [1.865] |
| Log Emolument | 9.234 [1.545] | 3.139 [0.353] | 9.017 [1.394] | 5.203 [0.391] |

Source: Authors' calculation based on ASI data. Note: The main entries and the brackets' entries are the mean and the standard deviation of each variable. We use the sample weights provided by ASI in the calculation.

## Results

This section presents the empirical evidence regarding the causal impact of the reforms in labor laws on various plant's employment and performances. We compare the difference in the outcome variables for the plants in Rajasthan relative to Punjab before and after the amendments. We then confirm that the reform had a differentially larger effect on the plants that are most likely to be "affected" than the "unaffected" using a triple difference framework. We estimate the triple difference specifications across the capital and labor-intensive industries and exporting and non-exporting industries to find the heterogeneous impact of the reform.

### Parallel Trends

We establish that the parallel trends assumption between the treatment and the control group holds using two approaches. First, we visually inspect the parallel trend assumption. Figure 1 provides the detrended values of employees, both directly employed as well as contractual workers for treated and the control states. Visually, the trends are parallel until 2014-15 and diverge thereafter. Second, we use a formal placebo regression, to check the potential treatment effects before the labor laws amendments (2011-12 to 2013-14).

$$Y_{ijst} = \beta_0\, Time_t + \beta_1\, Treat_i * Time_t + X_{it} + \kappa_i + \gamma_t + \delta_j t + \epsilon_{ijst} \qquad (3)$$

where $Time_t$ is a continuous variable (0,1,2) for the year 2011-12, 2012-13 and 2013-14 respectively. If the treatment and the control state holds parallel trends, then $\beta_1$ should be zero. In Table 4 and Table 5, we find the coefficient of the interaction term $Treat_i * Time_t$ to be statistically insignificant. Thus the parallel trend assumption holds and confirms that the results are not driven by spurious effects.

**Figure 1: Employment by treatment status**

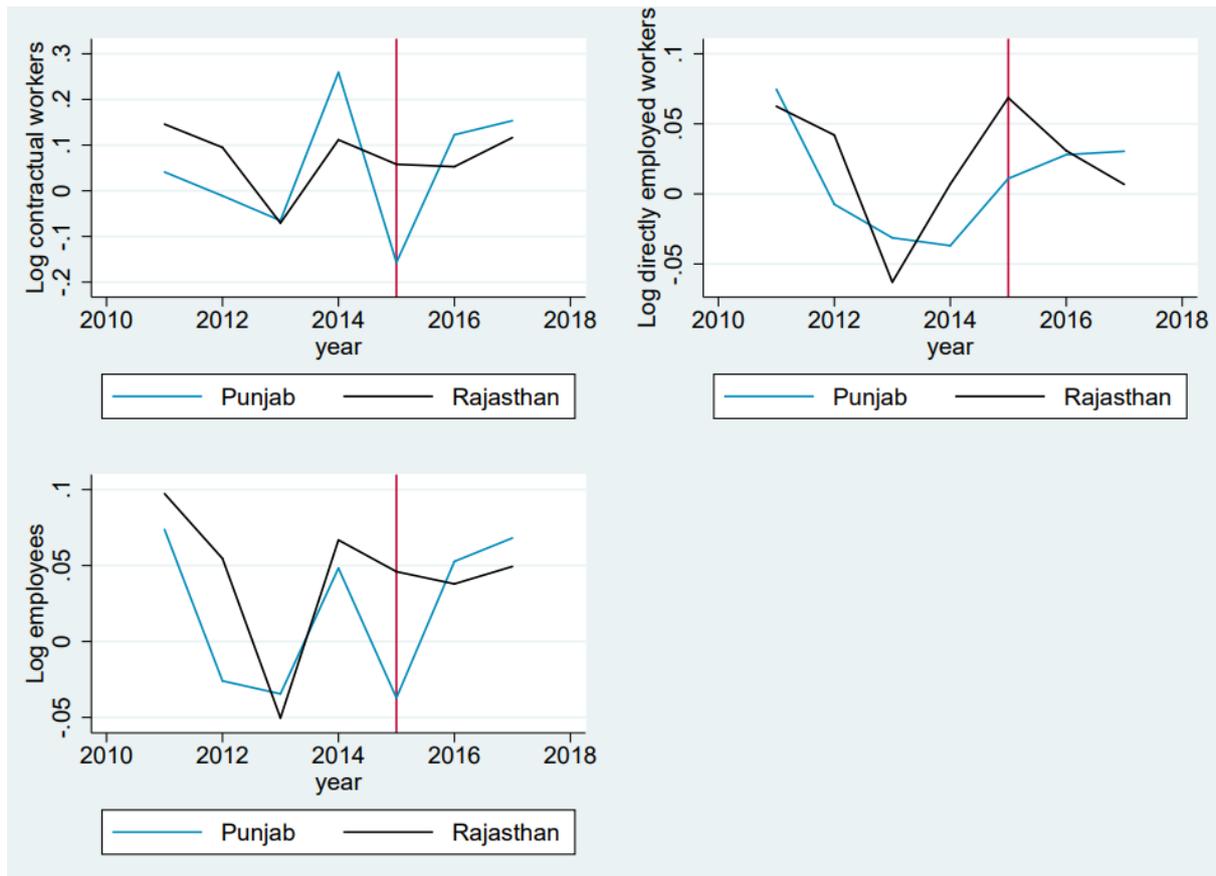

Source: Authors' calculation based on ASI data. Note: The values are detrended using first order differencing.

**Table 4: Testing the Parallel Trends for Plant Employment**

|  | Log Employees | Log Contractual Workers | Log Direct Workers | Contractual to Total Workers Ratio | Direct to Total Workers Ratio |
|---|---|---|---|---|---|
| ***Treat * Time*** | -0.008 | -0.036 | -0.020 | 0.001 | -0.002 |
|  | (0.004) | (0.007) | (0.005) | (0.000) | (0.002) |
| N | 9981 | 3043 | 8764 | 9914 | 8775 |
| r2 | 0.46 | 0.32 | 0.26 | 0.06 | 0.08 |
| Plant FE | Yes | Yes | Yes | Yes | Yes |
| Year FE | Yes | Yes | Yes | Yes | Yes |
| Industry Trends | Yes | Yes | Yes | Yes | Yes |
| Plant Controls | Yes | Yes | Yes | Yes | Yes |

Source: Authors' calculation based on ASI data. Note: Robust standard errors clustered at the state level in parentheses. All regressions include Treat and Time as control variables apart from various other plant controls. *** statistical significance at 1%; ** statistical significance at 5%; * statistical significance at 10%.

**Table 5: Testing the Parallel Trends for Plant Performances**

|  | Log Capital | Log Inputs | Log GVA | Log TFP (Method 1) | Log TFP (Method 2) | Log Profit | Log Emoluments |
|---|---|---|---|---|---|---|---|
| ***Treat* Time*** | 0.005 | -0.010 | -0.001 | -0.000 | -0.000 | -0.086 | -0.007 |
|  | (0.005) | (0.007) | (0.004) | (0.002) | (0.002) | (0.024) | (0.002) |
| N(shift bottom) | 9922 | 9922 | 9922 | 9917 | 9917 | 9930 | 9930 |
| r2 (capital) | 0.13 | 0.23 | 0.64 | 0.48 | 0.48 | 0.07 | 0.50 |
| Plant FE | Yes | Yes | Yes | Yes | Yes | Yes | Yes |
| Year FE | Yes | Yes | Yes | Yes | Yes | Yes | Yes |
| Industry Trends | Yes | Yes | Yes | Yes | Yes | Yes | Yes |
| Plant Controls | Yes | Yes | Yes | Yes | Yes | Yes | Yes |

Source: Authors' calculation based on ASI data. Note: Robust standard errors clustered at the state level in parentheses. All regressions include Treat and Time as control variables apart from various other plant controls. *** statistical significance at 1%; ** statistical significance at 5%; * statistical significance at 10%.

**Effect of Labor Law Amendments on Employment**

Table 6 presents the impact of the reforms in labor laws on employment outcomes. The outcome variables are log employees, log contractual workers, log direct workers, contractual to total workers, and direct to total workers. We find that the plants in Rajasthan differentially reduced total employees by around 3% compared to plants in Punjab, after relative to before the reforms. This suggests that flexibility in the labor laws resulted in the decline in plant employment. Furthermore, the plants responded to the reforms by decreasing the number of direct workers by around 2% whereas the decline in contract workers is insignificant. We notice that the ratio of both contractual to total workers and direct to total workers declined significantly by around 1% in Rajasthan after the reforms.

**Table 6: Effect on Employment**

| | Log Employees | Log Contractual Workers | Log Direct Workers | Contractual to Total Workers | Direct to Total Workers |
|---|---|---|---|---|---|
| *Post* Treat* | -0.029*** | 0.056 | -0.020** | -0.009*** | -0.005** |
| | (0.000) | (0.020) | (0.001) | (0.000) | (0.000) |
| N | 20128 | 6411 | 17579 | 20012 | 18021 |
| r2 | 0.51 | 0.27 | 0.26 | 0.05 | 0.07 |
| Plant FE | Yes | Yes | Yes | Yes | Yes |
| Year FE | Yes | Yes | Yes | Yes | Yes |
| Industry Trends | Yes | Yes | Yes | Yes | Yes |
| Plant Controls | Yes | Yes | Yes | Yes | Yes |

Source: Authors' calculation based on ASI data. Note: Robust standard errors clustered at the state level in parentheses. All regressions include Post and Treat as control variables apart from various other plant controls. *** statistical significance at 1%; ** statistical significance at 5%; * statistical significance at 10%.

The impact of the reforms is different for the plants that are most likely to be "affected" from the "unaffected". According to the Industrial Disputes Act (1947) reforms, we notice that the plants with workers greater than 100 and less than 300, and plants with less than or equal to 100 workers in 2014 are most likely to be affected. The triple difference estimates from Equation 2 in panel A of Table 7 indicate that the plants that had workers between 100 and 300 in 2014 experienced a significant increase in total employment and contractual workers, compared to the plants that had greater than 300 workers in Rajasthan compared to Punjab, in post compared to pre-treatment period. However, worryingly, these types of plants experienced a significant decline in direct to total workers. The plants with less than 100 workers in 2014 experienced a significant increase in total employment and direct employment compared to the plants with greater than 300 workers in Rajasthan compared to Punjab, in post compared to the pre-treatment period. Thus, the Industrial Disputes Act (1947) reforms successfully impacted those targeted plants.

According to the Contract Labor (Regulation and Abolition) Act (1970) reforms, we find that the plants with contractual workers greater than 20 and less than 50 in 2014 are most likely to be "affected". Consistent with the hypothesis, the affected plants experienced a significant increase in the ratio of contractual to total workers compared to the "unaffected" in Rajasthan compared to Punjab in post compared to the pre-treatment period (panel B of Table 7). Thus the reforms caused these "affected" plants to have a higher proportion of contractual workers. Also, these affected plants experienced a significant decline in both direct and contractual workers. Thus the effect on the "affected" plants is twofold i) experienced a decline in both contractual and direct workers ii) The proportion of contractual workers to total workers increased.

**Table 7: Heterogeneity in Employment for Affected Plants**

|  | Log Employees | Log Contractual Workers | Log Direct Workers | Contractual to Total Workers | Direct to Total Workers |
|---|---|---|---|---|---|
| *Panel A: Based on Changes in the Industrial Disputes Act (1947)* | | | | | |
| ***Treat* Post* 1. Affected*** | 0.061** | 0.055* | -0.004 | 0.015 | -0.024*** |
|  | (0.003) | (0.007) | (0.007) | (0.003) | (0.000) |
| ***Treat*Post*2.Affected*** | 0.037* | -0.002 | 0.016** | 0.005 | 0.004 |
|  | (0.438) | (0.695) | (0.590) | (0.082) | (0.074) |
| N | 20128 | 6411 | 17579 | 20012 | 18021 |
| r2 | 0.51 | 0.28 | 0.26 | 0.05 | 0.07 |
| Plant FE | Yes | Yes | Yes | Yes | Yes |
| Year FE | Yes | Yes | Yes | Yes | Yes |
| Industry Trends | Yes | Yes | Yes | Yes | Yes |
| Plant Controls | Yes | Yes | Yes | Yes | Yes |
| *Panel B: Based on Changes in the Contract Labor (Regulation and Abolition) Act (1970)* | | | | | |
| ***Treat*Post*Affected*** | -0.035 | -0.154** | -0.115*** | 0.034* | -0.008 |
|  | (0.006) | (0.008) | (0.000) | (0.005) | (0.005) |
| N | 20128 | 6411 | 17579 | 20012 | 18021 |
| r2 | 0.51 | 0.27 | 0.26 | 0.05 | 0.07 |
| Plant FE | Yes | Yes | Yes | Yes | Yes |
| Year FE | Yes | Yes | Yes | Yes | Yes |
| Industry Trends | Yes | Yes | Yes | Yes | Yes |
| Plant Controls | Yes | Yes | Yes | Yes | Yes |

Source: Authors' calculation based on ASI data. Note: Robust standard errors clustered at the state level in parentheses. "1.Affected" in Panel A are the plants with workers greater than 100 and less than 300, "2.Affected" in Panel A are the plants with less than equal to 100 workers, and the base is greater than 300 workers in 2014. All regressions in Panel A include Treat*Post, Post*Affected, Treat*Affected, Treat, Post, and Affected as control variables apart from various other plant controls. "Affected" in Panel B are the plants with contractual workers greater than 20 and less than 50 in 2014. All regressions in Panel B include Treat*Post, Post*Affected, Treat*Affected, Treat, Post, and Affected as control variables apart from various other plant controls. *** statistical significance at 1%; ** statistical significance at 5%; * statistical significance at 10%.

## Heterogeneity in Industry Characteristics

Table 8 presents the estimates on the reforms' heterogeneous impact based on industry characteristics. The results indicate that total employees and contractual to total employees declined whereas direct to total workers increased for the plants in labor-intensive industries compared to capital intensive industries in Rajasthan versus Punjab, and after compared to before the reform. These industries are preferring direct workers. This might be because of the plant-specific skill that the direct workers entail. Thus, the reforms helped the plants in the labor-intensive industries restructure the production factors according to their requirements.

As expected, plants in export-oriented industries experienced an increase in contractual to total workers than non-exporting industries in Rajasthan versus Punjab, and after compared to before the reform. The flexibility in the labor laws caused the plants in export-oriented industries to use contractual workers. The export market volatility causes this type of plant to choose contractual workers and thus reduce fixed cost of labor.

**Table 8: Heterogeneity in Employment based on Industry Characteristics**

|  | Log Employees | Log Contractual Workers | Log Direct Workers | Contractual to Total Workers | Direct to Total Workers |
|---|---|---|---|---|---|
| *Panel A: Labor Intensive Industries* | | | | | |
| *Treat*Post*LI* | -0.046* | -0.150 | 0.019 | -0.053** | 0.036* |
|  | (0.004) | (0.027) | (0.015) | (0.004) | (0.004) |
| N | 20128 | 6411 | 17579 | 20012 | 18021 |
| r2 | 0.51 | 0.27 | 0.26 | 0.05 | 0.07 |
| Plant FE | Yes | Yes | Yes | Yes | Yes |
| Year FE | Yes | Yes | Yes | Yes | Yes |
| Industry Trends | Yes | Yes | Yes | Yes | Yes |
| Plant Controls | Yes | Yes | Yes | Yes | Yes |
| *Panel B: Export Oriented Industries* | | | | | |
| *Treat*Post*EO* | -0.010 | 0.026 | -0.007 | 0.019*** | -0.003 |
|  | (0.007) | (0.052) | (0.012) | (0.000) | (0.003) |
| N | 20128 | 6411 | 17579 | 20012 | 18021 |
| r2 | 0.51 | 0.27 | 0.26 | 0.05 | 0.07 |
| Plant FE | Yes | Yes | Yes | Yes | Yes |
| Year FE | Yes | Yes | Yes | Yes | Yes |
| Industry Trends | Yes | Yes | Yes | Yes | Yes |
| Plant Controls | Yes | Yes | Yes | Yes | Yes |

Source: Authors' calculation based on ASI data. Note: Robust standard errors clustered at the state level in parentheses. LI= labor Intensive industries. All regression in Panel A includes Treat*Post, Post*LI, Treat*LI, Treat, Post, and LI as control variables apart from various other plant controls. EO= export-oriented industries. All regressions in Panel B include Treat*Post, Post*EO, Treat*EO, Treat, Post, and EO as control variables apart from various other plant controls. *** statistical significance at 1%; ** statistical significance at 5%; * statistical significance at 10%.

## Effect of Labor Law Amendments on Plant Performances and Inputs

We estimate Equation 1 to find the effect of deregulating the labor laws on various plant input and performances like capital, inputs, GVA, TFP, profits, and emoluments. The estimates are presented in Table 9. We do not find any significant impact of the labor laws on capital, inputs, profit, and emoluments. However, we find that the plants' productivity significantly increased by around 3% due to the increased flexibility in labor laws in Rajasthan compared to Punjab. We present the estimates of the impact of the the reform on plant inputs and performances by "affected" plants and industry heterogeneities in the Appendix.

**Table 9: Effect on Plant Performances and Inputs**

|  | Log Capital | Log Inputs | Log GVA | Log TFP (Method 1) | Log TFP (Method 2) | Log Profit | Log Emoluments |
|---|---|---|---|---|---|---|---|
| ***Post*Treat*** | -0.026 | -0.045 | 0.013* | 0.030* | 0.030* | 0.025 | -0.011 |
|  | (0.006) | (0.008) | (0.002) | (0.003) | (0.003) | (0.014) | (0.002) |
| N | 20020 | 20020 | 20020 | 20015 | 20015 | 20037 | 20037 |
| r2 | 0.10 | 0.28 | 0.77 | 0.70 | 0.70 | 0.38 | 0.55 |
| Plant FE | Yes | Yes | Yes | Yes | Yes | Yes | Yes |
| Year FE | Yes | Yes | Yes | Yes | Yes | Yes | Yes |
| Industry Trends | Yes | Yes | Yes | Yes | Yes | Yes | Yes |
| Plant Controls | Yes | Yes | Yes | Yes | Yes | Yes | Yes |

Source: Authors' calculation based on ASI data. Note: Robust standard errors clustered at the state level in parentheses. All regressions include Post and Treat as control variables apart from various other plant controls. *** statistical significance at 1%; ** statistical significance at 5%; * statistical significance at 10%.

**Plant Entry**

Is there more new plants' entry due to the increased flexibility in labor laws post 2014 in Rajasthan? To examine this change at the extensive margin, we estimate a separate regression at the 2-digit industry*state*year level. We identify a plant's birth from the "year of initial production" in the ASI data.

$$Entry_{jst} = \beta_0\ Treat_s + \beta_1\ Post_t + \beta_2\ Treat_s * Post_t + X_{jt} + \delta_j + \theta_t + \alpha_s + \omega_j t + \gamma_s t + \epsilon_{jst} \quad (5)$$

where $Entry_{jst}$ is the total entry of new plants in a year *t*, state *s,* and industry *j*. $Treat_s$ is an indicator variable that takes on the value 1 if the plant is in the treated state Rajasthan and 0 if the plant belongs to Punjab. $Post_t$ is a indicator variable that takes on the value 1 for the years after the amendments (2014-15 to 2016-17) and 0 otherwise. $\delta_j$ is the industry fixed effect that control for any time invariant unobserved heterogeneity at the industry level. $\theta_t$ is the year fixed effect and $\alpha_s$ is the state fixed effects. $\omega_j t$ is the industry trend and $\gamma_s t$ is the state trend. $X_{jt}$ are the control variables, namely, age of the plant, percentage of the output that the plant export, import dummy trend, gva,capital, inputs, profit, emoluments and workers. $\beta_2$ is the coefficient of interest that finds the impact of the reforms on the entry of new plants. We find from Table 10 that $\beta_2$ is statistically insignificant, and thus the impact of the reforms on plant employment and performances is from the incumbent plants in Rajasthan. We do not find the reforms to cause entry of new plants.

**Table 10: Effect on the number of plant entry**

|  | Plant Entry | Plant Entry |
|---|---|---|
| *Post*Treat* | -0.160 | -0.463 |
|  | (0.552) | (0.543) |
| N | 300 | 300 |
| r2 | 0.78 | 0.82 |
| Industry FE | Yes | Yes |
| Industry Trends | Yes | Yes |
| Year FE | Yes | Yes |
| State FE | Yes | Yes |
| State Trends | Yes | Yes |
| Control Variables | No | Yes |

Source: Authors' calculation based on ASI data. Note: This regression is estimated at a two-digit industry*state*year level. *** statistical significance at 1%; ** statistical significance at 5%; * statistical significance at 10%.

**Robustness Checks**

In this section, we perform several robustness checks of the main results presented above. For robustness, we use the establishments in the Indian states of Gujarat, Madhya Pradesh, Haryana, Uttar Pradesh as a control group (Panel A, B, C, D, and E of Table 11). We also estimate the DID in Equation 1 by making the control group as all the formal manufacturing establishments in India except those in Rajasthan (Panel F of Table 11). The results indicate that Rajasthan's labor reforms negatively impacted the total number of employees in an establishment. Moreover, this decline in employment is primarily through the decline in employment for direct workers. These results are qualitatively and quantitatively similar to the main results in Table 6, which indicate the labor laws' deregulations to cause a decline in employment.

We also limit the sample of establishments to the census sector and a balanced panel (Panel G and Panel H of Table 11). These establishments are larger, with greater than 100 employees. We donot notice a significant fall in the number of employees, but we notice a shift in the usage of directly employed workers to contractual workers. Worryingly, we find the labor deregulations in Rajasthan cause a decline in the direct workers and this has been offset by an increase in the contractual workers.

**Table 11: Effect of the labor reforms on employment for various specifications and various samples**

|  | Log Employees | Log Contractual Workers | Log Direct Workers | Contractual to Total Workers Ratio | Direct to Total Workers Ratio |
|---|---|---|---|---|---|
| *Panel A: Control group are the establishments in Gujarat* | | | | | |
| *Post*Treat* | -0.028** | -0.003 | -0.048* | 0.000 | -0.009 |
|  | (0.001) | (0.002) | (0.004) | (0.001) | (0.002) |
| N | 29236 | 10272 | 26801 | 29118 | 27196 |
| r2 | 0.50 | 0.27 | 0.20 | 0.05 | 0.07 |
| *Panel B: Control group are the establishments in Madhya Pradesh* | | | | | |
| *Post*Treat* | -0.067** | -0.052** | -0.088* | -0.004 | -0.003 |
|  | (0.002) | (0.001) | (0.011) | (0.004) | (0.000) |
| N | 15937 | 5177 | 14409 | 15823 | 14677 |
| r2 | 0.48 | 0.29 | 0.23 | 0.06 | 0.08 |
| *Panel C: Control group are the establishments in Haryana* | | | | | |
| Post*Treat | -0.082* | -0.057* | -0.092** | -0.027 | 0.006 |
| | (0.009) | (0.006) | (0.002) | (0.011) | (0.010) |
| N | 19651 | 7999 | 17184 | 19510 | 17651 |
| r2 | 0.48 | 0.25 | 0.19 | 0.05 | 0.06 |
| *Panel D: Control group are the establishments in Uttar Pradesh* | | | | | |
| Post*Treat | -0.024* | 0.063 | -0.077* | 0.007* | -0.023** |
| | (0.003) | (0.017) | (0.006) | (0.001) | (0.001) |
| N | 26608 | 9446 | 24425 | 26487 | 24791 |
| r2 | 0.50 | 0.23 | 0.19 | 0.05 | 0.06 |
| *Panel E: Control group are the establishments in all the neighboring states of Rajasthan* | | | | | |
| Post*Treat | -0.041** | 0.011 | -0.068*** | -0.004 | -0.010 |
| | (0.011) | (0.024) | (0.012) | (0.005) | (0.005) |
| N | 73328 | 26089 | 67122 | 73054 | 68204 |
| r2 | 0.50 | 0.23 | 0.19 | 0.03 | 0.05 |
| *Panel F: Control group are the establishments in all the Indian states except Rajasthan* | | | | | |
| Post*Treat | -0.042*** | -0.017 | -0.059*** | -0.005* | -0.006** |
| | (0.004) | (0.012) | (0.006) | (0.003) | (0.002) |
| N | 209474 | 74889 | 191922 | 208572 | 194956 |
| r2 | 0.49 | 0.21 | 0.20 | 0.02 | 0.04 |
| *Panel G: Establishments in the census sector* | | | | | |
| Post*Treat | -0.024 | 0.107** | -0.037* | -0.004 | -0.012* |
| | (0.006) | (0.005) | (0.004) | (0.001) | (0.002) |
| N | 10300 | 3737 | 9346 | 10259 | 9530 |
| r2 | 0.53 | 0.24 | 0.26 | 0.07 | 0.08 |
| *Panel H: Balanced panel* | | | | | |
| Post*Treat | -0.020 | 0.056* | -0.058** | 0.012* | -0.025* |
| | (0.003) | (0.005) | (0.003) | (0.002) | (0.002) |
| N | 5107 | 2102 | 4826 | 5104 | 4873 |
| r2 | 0.50 | 0.23 | 0.30 | 0.10 | 0.11 |

Source: Authors' calculation based on ASI data. Note: Robust standard errors clustered at the state level in parentheses. *** statistical significance at 1%; ** statistical significance at 5%; * statistical significance at 10%. We have controlled Post, Treat, plant fixed effects, year fixed effects, industry trends, and various other plant controls in all the regressions.

## Conclusion

In this paper, we empirically examined the impact of the 2014 labor laws deregulations in the Indian state of Rajasthan on plant employment and performances. The reform in the labor laws allowed us to utilize a quasi-natural experimental research design. We use a difference-in-difference specification to the establishment-level ASI panel data to examine the effects of Rajasthan's pro-employer reforms.

Our empirical analysis showed that the reforms had an unintended consequence of the decline in labor use. The implications regarding employment are similar to those presented by Roy, Dubey, and Ramaiah (2020), Deakin and Haldar (2015), Roychowdhury (2019a), Chatterjee and Kanbur (2015), D'Souza (2010), Kapoor (2014), Chandru(2014) in the sense that higher flexibility is associated with weaker employment growth. Also, worryingly, the increased flexibility resulted in the reduction in the directly employed workers. Heyes and Lewis (2015) and Avdagic (2015) find similar results for the European Union. If we consider plants as those that directly fall under the Industrial Disputes Act (1947) reforms, then we find these "affected" plants expand in labor use due to the reforms. If we consider the plants that fall directly under the Contract Labor (Regulation and Abolition) Act (1970) reforms, then we find that these "affected" plants experience greater use of contractual workers. We also find the reforms to cause the plants in the labor-intensive industries to restructure its production mix by reducing the labor use and preferring more direct workers. On the other hand, the labor laws' flexibility caused the plants in export-oriented industries to use more contractual workers. We also evaluate the reforms in labor laws on the plants' outcomes beyond the employment effects and find the reforms to positively impact GVA and productivity.